\documentclass[a4paper,11pt]{article}
\usepackage{pos}
\usepackage{cancel}

\title{Transfer matrices and temporal factorization of the Wilson fermion determinant}

\author*[a]{Urs Wenger}

\affiliation[a]{Albert Einstein Center for Fundamental Physics, Institute for Theoretical Physics, University of Bern, Sidlerstrasse 5, CH--3012 Bern, Switzerland}

\emailAdd{wenger@itp.unibe.ch}

\abstract{When lattice QCD is formulated in sectors of fixed quark numbers, the canonical fermion determinants can be expressed explicitly in terms of transfer matrices. This in turn provides a complete factorization of the fermion determinants in temporal direction. Here we present a generic overview of this factorization, apply it to Wilson-type fermions and provide explicit constructions of the transfer matrices. Possible applications of the factorization include multi-level integration schemes and the construction of improved estimators for generic $n$-point correlation functions.}

\FullConference{%
  The 39th International Symposium on Lattice Field Theory (Lattice2022),\\
  8-13 August, 2022 \\
  Bonn, Germany 
}

\DeclareMathOperator{\Tr}{Tr}

\newcommand{\minorindex}[2]{\, \backslash \hspace{-0.15cm} {#1}   \backslash  \hspace{-0.15cm}{#2}}

\def\D{{\cal D}}
\def\U{{\cal U}}

\def\T{{\cal T}}
\def\S{{\cal S}}

\def\Z{{\mathbb{Z}}}
\def\I{{\mathbb{I}}}

\begin{document}
\maketitle

\section{Introduction and motivation}
The temporal factorization of fermion determinants is a generic
feature of fermionic quantum field theories on the lattice. The key
steps for achieving the factorization are 1) the dimensional reduction
of the fermion determinant, 2) the projection to canonical sectors
with fixed fermion numbers, and 3) the factorization of the canonical
determinants in terms of transfer matrices. In these proceedings we
summarize these three steps first for a generic fermionic gauge field
theory and then exemplify the steps by applying them to QCD with
Wilson fermions.

The dimensional reduction of the fermion determinant has been known
since a long time
\cite{Blankenbecler:1981jt,Hirsch:1981jv,Wolff:1984ep} and has been
used in different contexts
\cite{Hasenfratz:1991ax,Adams:2004yy,deForcrand:2006zz,deForcrand:2007uz,Fodor:2007ga,Bilgici:2009gjc,Nagata:2010xi,Alexandru:2010yb,Giordano:2020roi}
since then. Canonical ensembles and determinants have found their use
in various applications \cite{Kratochvila:2005mk,
  Buhlmann:2021nsb}. In some cases they lead to a solution of the
fermion sign problem \cite{Wenger:2021qee}. For the factorization it
is important use the canonical projection of the fermion determinant
as first proposed \cite{Steinhauer:2014oda} and applied
\cite{Bergner:2015ywa,Bergner:2016qbz} in the context of
supersymmetric Yang-Mills quantum mechanics. Moreover, the
factorization of the fermion determinant has also been demonstrated
for the Hubbard model \cite{Burri:2019wge}.

The dimensional reduction and temporal factorization of the fermion
determinant is most interesting from an algorithmic point of
view. First, the dimensional reduction offers the possibility to
reduce the complexity of calculating the fermion determinant. Second,
the factorization allows the construction and application of
multi-level integration schemes following \cite{Ce:2016idq}. Since the
factorization in terms of the transfer matrices presented here is the
most atomic one, it may serve as the basis for more flexible and
efficient factorization schemes \cite{Ce:2016ajy,Giusti:2022xdh}. The
factorization also enables the construction of improved estimators for
generic $n$-point correlation functions
\cite{Burri:2019wge}. Moreover, the transfer matrices naturally
accommodate open boundary conditions in time.\\

To set the stage we consider a generic fermionic gauge field theory
with gauge fields $\U$ and fermion fields $\psi^\dagger, \psi$. Its
grand-canonical partition function at finite chemical potential $\mu$
is
\begin{align}
    Z_\text{GC}(\mu) &= \int \D\U \, e^{-S_b[\U]} \int {\cal D}\overline\psi {\cal D}\psi \, e^{-\overline\psi M[\U;\mu]\psi}\\
                     &=  \int \D\U\, e^{-S_b[\U]} \det M[\U;\mu] \, ,
                       \label{eq:Z_GC}
\end{align}
where $S_b[\U]$ is the bosonic gauge field action and $M[\U;\mu]$ the
fermion matrix. After integrating out the fermion fields one obtains
the fermion determinant $\det M[\U;\mu]$ as indicated in Eq.~(\ref{eq:Z_GC}).
 In general, this determinant is very difficult to calculate, due to its
 highly non-local dependence on the gauge field. In the Hamiltonian
 formulation one can formally write the partition function as a trace of the
 Hamiltonian Boltzman weight over all the states of the system, 
\begin{equation}
  Z_\text{GC}(\mu) = \Tr \left[e^{-{\cal H}(\mu)/T}\right] = \Tr
                     \prod_t \T_t(\mu)\, ,
\end{equation}
where the last equation indicates that on a space-time lattice the
temporal evolution can be written in terms of grand-canonical transfer
matrices $\T_t$ defined at fixed (Euclidean) times $t$. Finally, one
can use a fugacity expansion to relate the grand-canonical partition
function to the canonical one, $Z_C(N)$, for which the fermion number
$N$ is fixed,
\begin{equation}
    Z_\text{GC}(\mu)=\sum_{N} e^{-N \mu/T } \cdot Z_C(N) = \sum_{N}
    e^{-N \mu/T } \cdot \Tr \prod_t {\cal T}_t^{(N)} \, .
    \label{eq:relation Z_GC to Z_C}
\end{equation}
Here, ${\cal T}_t^{(N)}$ are the corresponding canonical transfer
matrices at fixed fermion number $N$. They can be obtained, at least
formally, by restricting $\T_t$ to states with fixed $N$. In the
following we show that these relations are not just formal, but can be
made explicit.

\section{Step 1: Dimensional reduction of the fermion determinant}
For generic gauge field theories discretized on a space-time lattice
with $L_s\times L_t$ lattice sites and a total of $L$ fermionic
degrees of freedom per time slice, the fermion matrix $M[\U;\mu]$ has
the (temporal) structure
\begin{equation}
M[\U;\mu] = \left(
\begin{array}{ccclc}
B_0 & e^{+\mu} C'_0 & 0 & \ldots & \pm e^{-\mu} C_{L_t-1} \\
e^{-\mu} C_0 & B_1 & e^{+\mu} C'_1 & & 0 \\
0   &e^{-\mu} C_1 & B_2 & \ddots &  \vdots\\
\vdots & \hfill\ddots & \hfill\ddots & & \\
&&&B_{L_t-2}& e^{+\mu} C'_{L_t-2} \\
\pm e^{+\mu} C'_{L_t-1}& 0 &  & e^{-\mu} C_{L_t-2} & B_{L_t-1}
\end{array} \right) \, .
\end{equation}
Here, the matrices $B_t$ describe the spatial fermion hoppings and
only depend on the spatial gauge fields at fixed time $t$, while the
matrices $C_t'$ and $C_t$ describe the temporal fermion hoppings
forward and backward in time, respectively. They only contain temporal
gauge fields. The $\pm$ signs in the upper right and lower left block
of the matrix indicate periodic or antiperiodic boundary conditions
for the fermions in the temporal direction.

Given this structure, the determinant of $M$ can be reduced by
iterative Schur decompositions yielding
\begin{equation}
\det M[\U;\mu] = \prod_{t=0}^{L_t-1} \det \tilde B_t \cdot \det\left(1
  \mp e^{\mu  L_t} \cdot {\cal T}\right) \, ,
\label{eq:reduced determinant}
\end{equation}
where ${\T = \T_0\cdot \ldots \cdot \T_{L_t-1}}$ with
$\T_t = \T_t[B_t,C_t,C_t']$, i.e., the matrices $\T_t$ only depend on
the spatial blocks associated with the time slice at time $t$. The
matrices $\tilde B_t$ are equal to $B_t$ up to constant factors of the
fugacity $e^{\pm \mu}$. Since the prefactor $\prod_t \det \tilde B_t$
is already factorized and hence not relevant in the remaining
derivation of the factorization, we neglect it for simplicity and only
reintroduce it at the very end.
 
The key object from step 1 is the matrix $\T[\U]$ given as the product
of spatial matrices,
\begin{equation}
  \T[\U] \equiv \prod_{t=0}^{L_t-1} \T_t \, .
\end{equation}
The matrices $\T_t$ are of size $L \times L$ only, while in contrast $M[\U;\mu]$ is
of size $(L\cdot L_t) \times (L\cdot L_t)$. 

\section{Step 2: Projection of the fermion determinant to canonical sectors}
The projection of the fermion determinant to canonical sectors
starts from the expansion of the fermion determinant in terms of the
fugacity $e^{-\mu/T}$,
\begin{equation}
\det M[\U;\mu] = \sum_{{N=-L/2}}^{L/2} e^{-{N}\cdot \mu/T } \cdot \det{}_{N}
M[\U] \, ,
\end{equation}
where $\det_{N} M[\U]$ denote the canonical determinants at fixed
fermion number $N$. Up to a constant multiplicative factor they are
simply given by the coefficients in the fugacity expansion of the
characteristic polynomial of the reduced matrix in
Eq.~(\ref{eq:reduced determinant}),
\begin{equation}
\sum_{{N=-L/2}}^{L/2} e^{-{N}\cdot \mu/T} \cdot \det{}_{N} M[\U] \propto \det\left(e^{-\mu/T} +   \T[\U]\right) \, ,
\end{equation}
where for simplicity we now restrict ourselves to antiperiodic
temporal boundary conditions for the fermions. The coefficients can be
calculated through the elementary symmetric functions $S_k$ of order
$k$ of the eigenvalues $\{\tau_i\}$ of $\T$,
\[
\det{}_N M[\U] \propto S_{L/2+N}(
\cal T) \, ,
\]
where
\begin{equation}
S_k(\T) \equiv S_k(\{\tau_i\}) = \sum_{1\leq
  i_1 < \cdots < i_k \leq L} \prod_{j=1}^k
\tau_{i_j} = \sum_{|J|=k} \det {\cal
  T}^{\, \backslash \hspace{-0.15cm} J \,
    \backslash  \hspace{-0.15cm}J} \, .
\label{eq:symmetric functions}
\end{equation}
In the last equality
we have made use of the fact that the symmetric functions $S_k$ can be
expressed in terms of the principal minors of order $k$ denoted by
$\det \T^{\minorindex{J}{J}}$. We recall that the principal minors are
obtained by computing the determinant of the matrix
$\T^{\minorindex{J}{J}}$ from which the columns and rows labeled by
the index set $J$ of size $k$ are removed.

Summarizing the derivation above we have
\begin{equation}
  \det{}_{{N}} M[\U] \propto \sum_J \det {\cal T}^{\, \backslash
    \hspace{-0.15cm} J  \, \backslash  \hspace{-0.15cm}J}[\U]  \propto \Tr
  \left[\prod_t {\cal T}_t^{({N})}\right]\, .
  \label{eq:canonical determinant}
\end{equation}
The last proportionality exposes the connection with the trace over
the fermionic states with fixed fermion number $N$ of the product of
transfer matrices in Eq.~(\ref{eq:relation Z_GC to Z_C}), except that
here the trace is taken for a fixed gauge field $\U$.

We note that the fermionic states are labeled by index sets
$J \subset \{1,\ldots,L\}, \, |J|={L/2+N}$, hence the number of states
is given by
\begin{equation}
  N_\text{states} = \left( 
\begin{array}{c}
L \\
{L/2+N}
\end{array}\right) = N_\text{principal minors} \, ,
\end{equation}
i.e., at half-filling ($N=0$) the number of states grows exponentially
with $L$. For relativistic gauge field theories, half-filling
corresponds to the vacuum sector where all states have equal numbers
of fermions and antifermions, hence $N=0$. At first sight, the
exponential growth of states looks like an obstacle for numerical
Monte-Carlo simulations, however, one can treat the index set $J$ as a
(discrete) dynamical degree of freedom which can be evaluated
stochastically
\cite{Steinhauer:2014oda,Bergner:2015ywa,Bergner:2016qbz,Burri:2019wge,Buhlmann:2021nsb}.

\section{Step 3: Temporal factorization of the fermion determinant}
Having the canonical fermion determinant at hand, we are now in the
position to derive its temporal factorization. To do so, we use the 
Cauchy-Binet formula
\begin{equation}
\det(A\cdot B)^{\minorindex{I}{K}} = \sum_{J}\det A^{\minorindex{I}{J}} \cdot \det B^{\minorindex{J}{K}}
\label{eq:Cauchy-Binet}
\end{equation}
to factorize the minor matrix of a product of matrices $A\cdot B$ into
the product of the corresponding minor matrices of $A$ and
$B$. Applying the Cauchy-Binet formula to the principal minors in
Eq.~(\ref{eq:canonical determinant}) achieves the
factorization. Reintroducing the prefactors $\det \tilde B_t$ and
defining
$(\T_t)_{IK} = \det \tilde B_t \cdot \det \T_t{}^{\minorindex{I}{K}}$
for simplicity we eventually obtain the expression in terms of the
transfer matrices,
\begin{equation}
 \det {\cal T}^{\minorindex{J}{J}}  = \det ({\T_0 \cdot\ldots\cdot
   \T_{L_t-1}})^{\minorindex{J}{J}}  = (\T_{0})_{JI} \cdot
 (\T_{1})_{IK} \cdot \ldots \cdot  (\T_{L_t-1})_{LJ} \, ,
\end{equation}
where implicit sums over the index sets $\{J, I, K, \ldots\}$ are
assumed.  Collecting everything we finally have
\begin{equation}
 \det{}_N M[\U] = \prod_{t=0}^{L_t-1} \det \tilde B_t \, 
 \cdot \sum_{\{J_t\}} \prod_{t=0}^{L_t-1} \det
 \T_t{}^{\minorindex{J_{t}}{J_{t+1}}} \, ,
\end{equation}
where $|J_t|={L/2+N}$ and $J_{L_t} \equiv J_{0}$.

\section{Application to QCD with Wilson fermions}
\label{sec:wilson fermion application}
We can now apply the three steps sketched in the previous sections to
QCD with Wilson fermions. For this purpose we consider the Wilson
fermion matrix $M[\U;\mu]$ for a single quark flavour with chemical
potential $\mu$,
\begin{equation*}
 M_\pm[\U;\mu] = \left(  
\begin{array}{ccccc}
 B_0  & P_+ A^+_0  &        &         & \pm P_- A^-_{L_t-1}\\
 P_- A^-_0 & B_1  & P_+ A^+_1   &        &  \\
      & P_- A^-_1  & B_2    & \ddots  &  \\
      &      & \ddots & \ddots &   \\
      &      &        &        &   P_+ A^+_{L_t-2} \\
\pm P_+ A^+_{L_t-1}& & &  P_- A^-_{L_t-2}    &  B_{L_t-1} \\
\end{array}
\right) \, 
\end{equation*}
with the Dirac projectors $P_\pm = \frac{1}{2}(\I \mp \Gamma_4)$.
Here, the temporal hoppings are
\[
A^+_t = e^{+\mu} \cdot \U_t = \left(A^-_t \right)^{-1} \quad
\text{with} \quad \U_t=\left\{{\mathbb{I}}_{4\times 4} \otimes
  U_4(\bar x, t), \, \bar x \in
\{0,\ldots, L_s^3-1\} \right\}
\]
collecting the temporal gauge links at fixed time $t$, while the
spatial fermion hoppings are collected in the spatial Wilson Dirac
operators $B_t$ containing only the spatial gauge links at time slice
$t$. All block matrices appearing in $M[\U;\mu]$ are
$(4\cdot N_c \cdot L_s^3 \times 4\cdot N_c\cdot L_s^3)$ matrices. The
reduced Wilson fermion determinant is then given by
\begin{equation}
\det M_{p,a}(\mu) \propto \prod_{t=0}^{L_t-1} \det Q_t^+ \cdot \det \left[ \I \pm
  e^{+\mu L_t} \T \right] \, ,
\label{eq:reduced Wilson determinant}
\end{equation}
where $\T$ is the product of spatial matrices 
\begin{equation}
  \T = \prod_{t=0}^{L_t-1} Q_t^+ \cdot \U_t \cdot \left(Q_{t+1}^-\right)^{-1}
  \equiv \prod_{t=0}^{L_t-1} \T_t
  \label{eq:Wilson transfer matrix product}
\end{equation}
with
\[
  Q_t^\pm = B_t P_\mp + P_\pm, \qquad
 B_t = \left(
      \begin{array}{cc}
        D_t & C_t \\
        -C_t & D_t
      \end{array}
    \right), 
\]
and hence
  \[  
     Q_t^+ = \left(
      \begin{array}{cc}
        1 & C_t \\
        0 & D_t
      \end{array}
    \right)    \, , \quad
     (Q_t^-)^{-1} = \left(
      \begin{array}{cc}
        D_t^{-1} & 0 \\
        C_t\cdot D_t^{-1}  & 1
      \end{array}
    \right)  \, .
  \]
We refer to Ref.~\cite{Alexandru:2010yb} for further details on the
derivation of the dimensional reduction for Wilson fermions.

The product of the spatial matrices  $Q^\pm_t$ and temporal gauge
links $\U_t$ can be written in different ways. The
form
\[
\T = \prod_t
       Q^+_t \cdot \U_{t} \cdot {(Q^-_{t+1})^{-1}}
\]
emphasises the connection to the usual definition of transfer matrices
between time slice at $t$ and $t+1$, while the form
\[
\T = \prod_t
\U^-_{t-1} \cdot (Q^-_t)^{-1} \cdot Q^+_t \cdot \U_{t}^+
\]
with $\U_t^\pm = \U_t P_\mp + P_\pm$ points out the separation of the
spatial gauge links within a fixed time slice $t$ contained in
$Q^\pm_t$ from those within neighbouring time slices at $t\pm
1$. There are also several ways to express the spatial matrices. The
form
\[
     \widetilde \T_t \equiv (Q_t^-)^{-1} \cdot Q_t^+ =
      \left(\begin{array}{cc}
              1   & 0 \\
              C_t & 1
            \end{array}\right)
      \left(\begin{array}{cc}
              D_t^{-1}   & 0 \\
              0         & D_t
            \end{array}\right)
      \left(\begin{array}{cc}
              1   & C_t \\
              0 & 1
            \end{array}\right)
        \]
exposes the relation  $\det \widetilde \T_t = 1$ and hence the
spectral property $\lambda \, \leftrightarrow \, 1/\lambda^*$ of the
eigenvalues $\lambda$ of $\widetilde \T_t$. In contrast, the form
\[      
      \widetilde \T_t =
      \left(\begin{array}{cc}
              D_t^{-1}   & D_t^{-1} \cdot C_t \\
              C_t \cdot  D_t^{-1} & D_t + C_t \cdot D_t^{-1} \cdot C_t
            \end{array}\right)
          \qquad \Leftrightarrow \qquad
          \widetilde \S_t =
      \left(\begin{array}{cc}
              C_t & D_t \\
              D_t & -C_t
            \end{array}\right) \, 
\]
expresses the relation between the matrix $\widetilde \T_t$
and the three-dimensional scattering matrix $\widetilde \S_t$. 

Given the reduced determinant in Eq.~(\ref{eq:reduced Wilson
  determinant}) and the explicit form of $\T$ in Eq.~(\ref{eq:Wilson
  transfer matrix product}) it is now straighforward to apply step 2
for the case of Wilson fermions and project to the canonical
determinants with $N_q$ quarks,
\begin{equation}
\det M_{N_q} = \prod_t \det Q_t^+ \cdot \sum_{A} \det \T^{\bcancel A
  \bcancel A} \, .
\label{eq:canonical QCD determinant}
\end{equation}
Here, the sum is over all index sets
$A \subset \{1,2,\ldots, 2 N_q^\text{max}\}$ of size
$|A| = N_q^\text{max} + N_q$ where
$N_q^\text{max} = 2\cdot N_c\cdot L_s^3$ for gauge group SU($N_c$) and
$N_q \in \{-N_c\cdot L_s^3, \ldots, +N_c\cdot L_s^3\}$. Finally, the
temporal factorization of the QCD determinant is achieved by applying
step 3 to Eq.~(\ref{eq:canonical QCD determinant}) yielding
\begin{equation}
\det M_{N_q} = \prod_t \det Q_t^+ \cdot \prod_t M\left(\left(Q_t^-\right)^{-1}\right)_{\bcancel{A_t}\bcancel{B}_t}
      M( Q_t^+)_{\bcancel{B}_t\bcancel{C}_t}
      M(\U_t)_{\bcancel{C}_t\bcancel{A_{\hspace{0.1cm}}}\hspace{-0.1cm}_{t+1}}
      \label{eq:factorized QCD determinant}
\end{equation}
where for notational simplicity we have introduced the notation $M(A)$
for the minor matrix of a generic matrix $A$.

We end this section by pointing out three interesting properties of the
minor matrices appearing in Eq.~(\ref{eq:factorized QCD
  determinant}). First, we note that transforming all temporal gauge
links at one fixed time slice $t$ with an element $z_k = e^{2\pi i
  \cdot k/N_c} \in \Z(N_c)$ of the center of the gauge group, i.e.,
\[
  \U_{t} \rightarrow \U_{t}' =  z_k\cdot \U_{t}
\]
we find
\[
  \det M_{N_q} \rightarrow  \det M_{N_q}' = \prod_t \det Q_t^+ \cdot \sum_{A} \det (z_k\cdot\T)^{\bcancel A \bcancel A} 
= z_k^{- N_q} \cdot \det M_{N_q} \, .
\]
As a consequence, summing over $z_k, k=1,\ldots,N_c$ yields
\[
\det M_{N_q} = 0  \qquad \text{for} \quad \, N_q \neq 0 \,\text{mod}\, N_c
\, ,
\]
i.e., only sectors with integer baryon numbers yield nonvanishing
canonical partition functions. This nontrivial physical relation
between the quark and baryon numbers in QCD becomes trivial in the
factorized canonical formulation. Second, we note that with the relations
\[
  M(Q^{-1})_{\bcancel{A_{}}\bcancel{B}} = (-1)^{p(A,B)} \frac{\widetilde
    M(Q)_{BA}}{\det Q} \, , \quad \det Q^+_t = \det Q^-_t \, ,
\]
where $\widetilde M(Q)$ is the complementary minor matrix of $Q$ and
$p(A,B)$ the total parity of the index sets $A$ and $B$, the inversion
of $Q_t^{-}$ can be avoided. Third, we note that $\U_t$ is
  trivial in Dirac space and has a simple block structure in terms of
  the collection of temporal gauge links
  $W_t = {\mathbb{I}}_{4\times 4} \otimes U_4(\bar x, t)$ at fixed
spatial site $\bar x$. As a consequence, the corresponding minor
matrix element
$M(\U_t)_{\bcancel{C}_t\bcancel{A_{\hspace{0.1cm}}}\hspace{-0.1cm}_{t+1}}$
in Eq.~(\ref{eq:factorized QCD determinant})
  is nonzero only if
  $M(W_t)_{\bcancel c_t \bcancel a_{t+1}} \neq 0, \forall \bar x$,
  where $c_t(\bar x) \in C_t$ and $a_{t+1}(\bar x) \in A_{t+1}$ are
  the sub-index sets restricted to $\bar x$. Hence,
  $M(\U_t)_{\bcancel{C}_t\bcancel{A_{\hspace{0.1cm}}}\hspace{-0.1cm}_{t+1}}
    = 0$ if $|c_t(\bar x)| \neq |a_{t+1}(\bar x)|$ at any of the sites
    $\bar x$. This imposes a considerable restriction on the allowed
    index sets.

\section{Multi-level integration schemes and improved estimators}
The factorization provided by Eq.~(\ref{eq:factorized QCD
  determinant}) allows for simple multi-level integration schemes,
since the gauge fields on different time slices are no longer coupled
through the fermion determinant. For example, the temporal gauge links
$\U_t$ at different times $t$ are completely decoupled from each
other. Since the spatial matrix $\U_t$ is block-diagonal (see remark
above), $M(\U_t)$ is trivial to calculate. Assuming Wilson's plaquette
gauge action the only interaction
between the temporal gauge links at fixed $t$ is through the temporal
plaquettes only.

The spatial gauge fields at fixed time $t$ on the other hand interact
with each other only through $M\left((Q^-_t)^{-1}\right)$ and
$M(Q^+_t)$.
Of course the spatial gauge links at two neighbouring time slices are
still coupled through the gauge action, however, assuming again
Wilson's plaquette gauge action, the coupling is through the temporal
plaquettes only.

The caveat for practical implementations of such multi-level
integration schemes lies in the fact that a priori the factors in
Eq.~(\ref{eq:factorized QCD determinant}) are not necessarily
positive. How severe the corresponding sign problem is and whether it
can be ameliorated by multi-level integration schemes remains to be seen.

For the construction of fermionic observables, such as $n$-point
correlation functions, we follow Ref.~\cite{Burri:2019wge} where the
construction is described using the Hubbard model as an
example. Source and sink operators $\S$ and $\overline \S$,
respectively, inserted at time $t$ simply remove or add indices
from/to the index sets at time $t$. The operators $\S$ and
$\overline \S$ potentially change the quark number $N_q$ and hence the
canonical sector, e.g.,
\begin{equation}
      \ldots \cdot \T_{t-1}^{(N_q)} \cdot \S_{N_q\rightarrow N_q+3} \cdot \T_{t}^{(N_q+3)}\cdot  \ldots \cdot \T_{t'}^{(N_q+3)}
      \cdot \overline \S_{N_q+3\rightarrow N_q} \cdot
      \T_{t'+1}^{(N_q)}\cdot \ldots \, .
      \label{eq:correlation function}
\end{equation}
Starting from, e.g., the vacuum sector with $N_q=0$ the example in
Eq.~(\ref{eq:correlation function}) corresponds to a baryon-antibaryon
correlation function $C_{\text{B}-\overline{\text{B}}}(t'-t)$.

In the factorized formulation, it is natural to construct improved
estimators as follows. Barring potential sign problems one can
directly simulate a correlation function $C(t'-t)$, e.g., as given in
Eq.~(\ref{eq:correlation function}),  at large $t'-t$ and determine $C(t'+1-t)$
relative to $C(t'-t)$ through the expectation value
\[
  \langle C(t'+1-t)\rangle_{C(t'-t)} \sim e^{- a E} \, .
\]
That is, the ground-state energy $E$ of the correlator is essentially
determined by measuring the effect of shifting
$\overline \S_{N_q+3\rightarrow N_q}$ from $t'$ to $t'+1$ and changing
$ \T_{t'+1}^{(N_q)} \rightarrow \T_{t'+1}^{(N_q+3)}$ on top of the
correlation function $C(t'-t)$. If this can be implemented in
practice, it would open the way to tackle signal-to-noise problems on
top of using multi-level integration schemes.

The construction of improved estimators for correlation functions is
closely related to the expectation value of the transfer matrix
$\langle \T_t^{(N_q)} \rangle_{N_q}$. In principle, this object
contains all the spectral information of the system in the canonical
sector with $N_q$ quarks, but in practice it is difficult to calculate
since the size of the transfer matrix grows exponentially with the
spatial lattice size $L_s$. Nevertheless, since the theory is local,
one can expect that only a limited number of matrix elements are
necessary to approximate the low-lying spectrum of the transfer
matrix.

\section{Summary and outlook}
In these proceedings we have summarized the generic steps leading to a
complete temporal factorization of the fermion determinant for generic
fermionic gauge field theories. The steps involve 1) the dimensional
reduction of the fermion determinant, 2) the projection to canonical
sectors with fixed fermion numbers, and 3) the temporal factorization
in terms of transfer matrices. Applying these three steps to QCD with
Wilson fermions leads to the most atomic temporal factorization of the
Wilson fermion determinant as given in Eq.~(\ref{eq:factorized QCD
  determinant}). The factorization opens the way for more flexible and
potentially more efficient multi-level integration schemes for
QCD. The main caveat for making further progress in this direction
lies in the fact that the factors in Eq.~(\ref{eq:factorized QCD
  determinant}) are a priori not necessarily positive and hence may
induce a potential sign problem. However, it is worthwhile to point
out that the matrices $Q^{\pm}$ are strictly positive, and hence also
all their principal minors.

The generic canonical projection and subsequent factorization outlined
here has already been applied successfully to a range of fermionic
(gauge) field theories in various computational setups. In
\cite{Steinhauer:2014oda,Bergner:2015ywa, Bergner:2016qbz} the
principal minors of the canonical projection have been simulated in
one-dimensional supersymmetric SU($N_c$) Yang-Mills gauge theories. In
\cite{Burri:2019wge} it was demonstrated in the Hubbard model that the
Hubbard-Stratanovich field can be analytically integrated out from the
factorized determinant such that the model can be simulated with the
discrete index sets (representing the fermion occupation numbers) as
the only remaining degrees of freedom. In low dimensions, the
positivity of the fermion weights can then be proven for any arbitrary
spin- and mass-imbalanced system. In \cite{Wenger:2021qee} the
dimensional reduction and determinant factorization has been used to
derive the exact three-dimensional effective Polyakov-loop action for
QCD in the heavy-dense limit. Similar to the Hubbard model, the
temporal gauge fields can be integrated out analytically leading to a
system which is free of the fermion sign problem at finite baryon
density. In \cite{Buhlmann:2021nsb} we reported on the canonical
projection of the Wilson fermion determinant for the case of the
two-flavour Schwinger model. The canonical fermion determinants can
then be used, e.g., to calculate meson-scattering phase shifts from
finite-volume effects. A summary of the results of some of these
applications is currently in preparation.

The temporal factorization presented here is loosely related to other
approaches to determinant factorizations. For example, it is probably
straightforward to derive the factorization based on winding number
expansion techniques \cite{Danzer:2008xs} starting from the transfer
matrices derived here. Furthermore, fermion bags similar to the ones
introduced in \cite{Chandrasekharan:2009wc} can be identified
straightforwardly using the index sets. In our approach, the bags are
confined to time slices at fixed $t$ (with weights given by the minor
matrices $M(Q^\pm_t)$), however, the bags can be naturally extended in
time by connecting the index sets in time. Following this line of
thought a little further immediately leads to the interpretation of
the index sets as fermion occupation numbers and fermion loops as
suggested in \cite{Steinhauer:2014oda}.  Finally, we note that the
factorization of the Wilson fermion determinant presented in
Sec.~\ref{sec:wilson fermion application} is closely related to the
construction of the transfer matrices in
\cite{Luscher:1976ms}. However, we have so far not established the
exact relation between the two
constructions.\\

\noindent {\bf Acknowledgements:} I would like to thank Patrick
B\"uhlmann for useful discussions.

\bibliographystyle{JHEP}
\bibliography{tmatfftwfd_PoS}

\providecommand{\href}[2]{#2}\begingroup\raggedright\begin{thebibliography}{10}

\bibitem{Blankenbecler:1981jt}
R.~Blankenbecler, D.~J. Scalapino and R.~L. Sugar, \emph{{Monte Carlo
  Calculations of Coupled Boson - Fermion Systems. 1.}},
  \href{http://dx.doi.org/10.1103/PhysRevD.24.2278}{\emph{Phys. Rev. D} {\bf
  24} (1981) 2278}.

\bibitem{Hirsch:1981jv}
J.~E. Hirsch, D.~J. Scalapino, R.~L. Sugar and R.~Blankenbecler, \emph{{An
  Efficient Monte Carlo Procedure for Systems With Fermions}},
  \href{http://dx.doi.org/10.1103/PhysRevLett.47.1628}{\emph{Phys. Rev. Lett.}
  {\bf 47} (1981) 1628--1631}.

\bibitem{Wolff:1984ep}
U.~Wolff, \emph{{A reduction of the kogut-susskind fermion determinant}},
  \href{http://dx.doi.org/10.1103/PhysRevD.30.2236}{\emph{Phys. Rev. D} {\bf
  30} (1984) 2236}.

\bibitem{Hasenfratz:1991ax}
A.~Hasenfratz and D.~Toussaint, \emph{{Canonical ensembles and nonzero density
  quantum chromodynamics}},
  \href{http://dx.doi.org/10.1016/0550-3213(92)90247-9}{\emph{Nucl. Phys. B}
  {\bf 371} (1992) 539--549}.

\bibitem{Adams:2004yy}
D.~H. Adams, \emph{{A Dimensionally reduced expression for the QCD fermion
  determinant at finite temperature and chemical potential}},
  \href{http://dx.doi.org/10.1103/PhysRevD.70.045002}{\emph{Phys. Rev. D} {\bf
  70} (2004) 045002}, [\href{http://arxiv.org/abs/hep-th/0401132}{{\tt
  hep-th/0401132}}].

\bibitem{deForcrand:2006zz}
P.~de~Forcrand and U.~Wenger, \emph{{New baryon matter in the lattice
  Gross-Neveu model}}, \href{http://dx.doi.org/10.22323/1.032.0152}{\emph{PoS}
  {\bf LAT2006} (2006) 152}, [\href{http://arxiv.org/abs/hep-lat/0610117}{{\tt
  hep-lat/0610117}}].

\bibitem{deForcrand:2007uz}
P.~de~Forcrand, M.~A. Stephanov and U.~Wenger, \emph{{On the phase diagram of
  QCD at finite isospin density}}, {\emph{PoS} {\bf LATTICE2007} (2007) 237},
  [\href{http://arxiv.org/abs/0711.0023}{{\tt 0711.0023}}].

\bibitem{Fodor:2007ga}
Z.~Fodor, K.~K. Szabo and B.~C. Toth, \emph{{Hadron spectroscopy from canonical
  partition functions}},
  \href{http://dx.doi.org/10.1088/1126-6708/2007/08/092}{\emph{JHEP} {\bf 08}
  (2007) 092}, [\href{http://arxiv.org/abs/0704.2382}{{\tt 0704.2382}}].

\bibitem{Bilgici:2009gjc}
E.~Bilgici, J.~Danzer, C.~Gattringer, C.~B. Lang and L.~Liptak,
  \emph{{Canonical fermion determinants in lattice QCD: Numerical evaluation
  and properties}},
  \href{http://dx.doi.org/10.1016/j.physletb.2011.01.035}{\emph{Phys. Lett. B}
  {\bf 697} (2011) 85--89}, [\href{http://arxiv.org/abs/0906.1088}{{\tt
  0906.1088}}].

\bibitem{Nagata:2010xi}
K.~Nagata and A.~Nakamura, \emph{{Wilson Fermion Determinant in Lattice QCD}},
  \href{http://dx.doi.org/10.1103/PhysRevD.82.094027}{\emph{Phys. Rev. D} {\bf
  82} (2010) 094027}, [\href{http://arxiv.org/abs/1009.2149}{{\tt 1009.2149}}].

\bibitem{Alexandru:2010yb}
A.~Alexandru and U.~Wenger, \emph{{QCD at non-zero density and canonical
  partition functions with Wilson fermions}},
  \href{http://dx.doi.org/10.1103/PhysRevD.83.034502}{\emph{Phys. Rev. D} {\bf
  83} (2011) 034502}, [\href{http://arxiv.org/abs/1009.2197}{{\tt 1009.2197}}].

\bibitem{Giordano:2020roi}
M.~Giordano, K.~Kapas, S.~D. Katz, D.~Nogradi and A.~Pasztor, \emph{{New
  approach to lattice QCD at finite density; results for the critical end point
  on coarse lattices}},
  \href{http://dx.doi.org/10.1007/JHEP05(2020)088}{\emph{JHEP} {\bf 05} (2020)
  088}, [\href{http://arxiv.org/abs/2004.10800}{{\tt 2004.10800}}].

\bibitem{Kratochvila:2005mk}
S.~Kratochvila and P.~de~Forcrand, \emph{{The Canonical approach to finite
  density QCD}}, \href{http://dx.doi.org/10.22323/1.020.0167}{\emph{PoS} {\bf
  LAT2005} (2006) 167}, [\href{http://arxiv.org/abs/hep-lat/0509143}{{\tt
  hep-lat/0509143}}].

\bibitem{Buhlmann:2021nsb}
P.~B\"uhlmann and U.~Wenger, \emph{{Finite-volume effects and meson scattering
  in the 2-flavour Schwinger model}},
  \href{http://dx.doi.org/10.22323/1.396.0463}{\emph{PoS} {\bf LATTICE2021}
  (2022) 463}, [\href{http://arxiv.org/abs/2112.15228}{{\tt 2112.15228}}].

\bibitem{Wenger:2021qee}
U.~Wenger and P.~B\"uhlmann, \emph{{Heavy-dense QCD at fixed baryon number
  without a sign problem}},
  \href{http://dx.doi.org/10.22323/1.396.0584}{\emph{PoS} {\bf LATTICE2021}
  (2022) 584}, [\href{http://arxiv.org/abs/2110.15021}{{\tt 2110.15021}}].

\bibitem{Steinhauer:2014oda}
K.~Steinhauer and U.~Wenger, \emph{{Loop formulation of supersymmetric
  Yang-Mills quantum mechanics}},
  \href{http://dx.doi.org/10.1007/JHEP12(2014)044}{\emph{JHEP} {\bf 12} (2014)
  044}, [\href{http://arxiv.org/abs/1410.0235}{{\tt 1410.0235}}].

\bibitem{Bergner:2015ywa}
G.~Bergner, H.~Liu and U.~Wenger, \emph{{Canonical simulations of
  supersymmetric SU(N) Yang-Mills quantum mechanics}},
  \href{http://dx.doi.org/10.22323/1.251.0241}{\emph{PoS} {\bf LATTICE2015}
  (2016) 241}, [\href{http://arxiv.org/abs/1509.01446}{{\tt 1509.01446}}].

\bibitem{Bergner:2016qbz}
G.~Bergner, H.~Liu and U.~Wenger, \emph{{A local update algorithm for
  supersymmetric Yang-Mills quantum mechanics}},
  \href{http://dx.doi.org/10.22323/1.256.0395}{\emph{PoS} {\bf LATTICE2016}
  (2016) 395}, [\href{http://arxiv.org/abs/1612.04291}{{\tt 1612.04291}}].

\bibitem{Burri:2019wge}
S.~Burri and U.~Wenger, \emph{{The Hubbard model in the canonical
  formulation}}, \href{http://dx.doi.org/10.22323/1.363.0249}{\emph{PoS} {\bf
  LATTICE2019} (2019) 249}, [\href{http://arxiv.org/abs/1912.09361}{{\tt
  1912.09361}}].

\bibitem{Ce:2016idq}
M.~C\`e, L.~Giusti and S.~Schaefer, \emph{{Domain decomposition, multi-level
  integration and exponential noise reduction in lattice QCD}},
  \href{http://dx.doi.org/10.1103/PhysRevD.93.094507}{\emph{Phys. Rev. D} {\bf
  93} (2016) 094507}, [\href{http://arxiv.org/abs/1601.04587}{{\tt
  1601.04587}}].

\bibitem{Ce:2016ajy}
M.~C\`e, L.~Giusti and S.~Schaefer, \emph{{A local factorization of the fermion
  determinant in lattice QCD}},
  \href{http://dx.doi.org/10.1103/PhysRevD.95.034503}{\emph{Phys. Rev. D} {\bf
  95} (2017) 034503}, [\href{http://arxiv.org/abs/1609.02419}{{\tt
  1609.02419}}].

\bibitem{Giusti:2022xdh}
L.~Giusti and M.~Saccardi, \emph{{Four-dimensional factorization of the fermion
  determinant in lattice QCD}},
  \href{http://dx.doi.org/10.1016/j.physletb.2022.137103}{\emph{Phys. Lett. B}
  {\bf 829} (2022) 137103}, [\href{http://arxiv.org/abs/2203.02247}{{\tt
  2203.02247}}].

\bibitem{Danzer:2008xs}
J.~Danzer and C.~Gattringer, \emph{{Winding expansion techniques for lattice
  QCD with chemical potential}},
  \href{http://dx.doi.org/10.1103/PhysRevD.78.114506}{\emph{Phys. Rev. D} {\bf
  78} (2008) 114506}, [\href{http://arxiv.org/abs/0809.2736}{{\tt 0809.2736}}].

\bibitem{Chandrasekharan:2009wc}
S.~Chandrasekharan, \emph{{The Fermion bag approach to lattice field
  theories}}, \href{http://dx.doi.org/10.1103/PhysRevD.82.025007}{\emph{Phys.
  Rev. D} {\bf 82} (2010) 025007}, [\href{http://arxiv.org/abs/0910.5736}{{\tt
  0910.5736}}].

\bibitem{Luscher:1976ms}
M.~L{\"u}scher, \emph{{Construction of a Selfadjoint, Strictly Positive
  Transfer Matrix for Euclidean Lattice Gauge Theories}},
  \href{http://dx.doi.org/10.1007/BF01614090}{\emph{Commun. Math. Phys.} {\bf
  54} (1977) 283}.

\end{thebibliography}\endgroup

\end{document}